\documentstyle[11pt,cosmos,epsfig]{article}

\def\gev{\mbox{GeV}}
\def\mev{\mbox{MeV}}

\def\cm{\mbox{cm}}
\def\mpc{\mbox{Mpc}}
 
\def\AJ{{\it Ap. J.} }
\def\AJL{{\it Ap. J. Lett.} }

\def\APJ{{\it Ap. J.} }

\def\IJMP{{\it Int. J. Mod. Phys.} }

\def\MNRAS{{\it Mon. Not. R. Ast. Soc.} }
\def\NAT{{\it Nature} }
\def\NAST{{\it New Astronamy}}

\def\PL{{\it Phys. Lett.} }
\def\PR{{\it Phys. Rev.} }
\def\PRL{{\it Phys. Rev. Lett.} }

\def\SC{{\it Science} }

\def\frac#1#2{{\textstyle{{#1}\over {#2}}}}

\def\lsim{\mathrel{\rlap{\lower4pt\hbox{\hskip1pt$\sim$}}
    \raise1pt\hbox{$<$}}}
\def\gsim{\mathrel{\rlap{\lower4pt\hbox{\hskip1pt$\sim$}}
    \raise1pt\hbox{$>$}}}
\def\sqr#1#2{{\vcenter{\vbox{\hrule height.#2pt
         \hbox{\vrule width.#2pt height#1pt \kern#1pt
         \vrule width.#2pt}
         \hrule height.#2pt}}}}

\newcommand{\beq}{\begin{equation}}
\newcommand{\eeq}{\end{equation}}
\newcommand{\bea}{\begin{eqnarray}}
\newcommand{\eea}{\end{eqnarray}}

\begin{document}

\title{Self-Interacting Scalar Dark Matter and Higgs Decay }

\author{M.C. Bento, O. Bertolami}
\affil{Instituto Superior T\'ecnico, Departamento de F\'\i sica,
Av. Rovisco Pais 1, 1049-001 Lisboa, Portugal }
\author{\underline{R. Rosenfeld}}
\affil{Instituto de F\'\i sica Te\'orica,
R.\ Pamplona 145, 01405-900 S\~ao Paulo - SP, Brazil}
\author{L. Teodoro}
\affil{Centro Multidisciplinar de Astrof\'\i sica, 
Instituto Superior T\'ecnico,
Av. Rovisco Pais 1, 1049-001 Lisboa, Portugal}

\begin{abstract}
Self-interacting dark matter has been suggested in order to overcome the
difficulties of the Cold Dark Matter model on galactic scales. We argue that
a scalar gauge singlet coupled to the Higgs boson, leading to an invisibly 
decaying Higgs, is an interesting candidate for this 
self-interacting dark matter particle.
\end{abstract}


\section{Introduction}

Finding clues for the nature of dark matter in the Universe is one of the
most
pressing issues in the interface between particle physics and cosmology.
In this talk we briefly review some work we have done [\cite{us}] 
in the direction
of finding a suitable, particle physics motivated, candidate that could solve 
some of the recent problems with the usual cold dark matter scenario. 

The cold dark matter model (CDM) supplemented by a cosmological constant 
sucessfully explains, in
 the context of inflationary models, the observed structure of the 
Universe on large scales,
the cosmic microwave background anisotropies and  type Ia 
supernovae observations [\cite{Bahcall}] for a given set of density 
parameters, {\it e.g.},  $\Omega_{DM} \sim 0.30$, 
$\Omega_{Baryons} \sim 0.05$ and $\Omega_{\Lambda} \sim 0.65$. 
According to this scenario, initial
Gaussian density fluctuations, mostly in non-relativistic collisionless
particles, the so-called cold dark matter, are generated in an inflationary 
period of the Universe. These fluctuations grow gravitationally forming dark
halos into which luminous matter is eventually condensed and cooled. 

However, despite its successes, there is a growing wealth
of recent observational data that raise problems in
the CDM scenarios. N-body simulations  
predict a number of halos which is a factor $\sim$ 10 larger than the observed
number at the 
level of Local Group [\cite{Mooreetal2,Klypin}]. Furthermore, CDM models 
yield dispersion velocities in the Hubble flow within a sphere of 
$5~h^{-1}~$Mpc
between $300-700~$kms$^{-1}$ for $\Omega_{DM} \sim 0.95$ and 
between $150-300~$kms$^{-1}$  for $\Omega_{DM} \sim 0.30$. 
The observed value is about 
$60~$kms$^{-1}$. Neither model 
can produce a single  Local Group candidate with the observed velocity 
dispersion in a simulation box of comoving volume 
$10^{6}~h^{-3}\mbox{Mpc}^{3}$ [\cite{Governato}]. A 
related issue is that astrophysical systems which are DM dominated like 
dwarf galaxies [\cite{Moore,Flores-Primack,Burkert1}], low surface brightness
galaxies
[\cite{Blok}] and galaxy clusters without a central cD galaxy [\cite{Tyson}]
show shallow matter--density profiles which can be modeled by 
isothermal spheres with finite central densities.
This is in contrast with galactic and galaxy cluster halos in high 
resolution N-body simulations 
[\cite{Navarro,Ghigna0,Ghigna,Mooreetal}] which have singular cores, 
with $\rho \sim r^{-\gamma}$ and $\gamma$ in the range between 1 and 2.
Indeed, cold collisionless DM particles do not have 
any associated length scale leading, due to hierarchical gravitational 
collapse, to dense dark matter halos with negligible 
core radius [\cite{Hogan}].

A possible solution, coming from particle physics, would be to 
allow DM particles to self-interact so that they have a large scattering 
cross section and negligible annihilation or dissipation. Self-interaction
induces  a characteristic length scale via  the mean free 
path of the particle in the halo.
This idea has been originally proposed to suppress small scale power 
in the standard CDM model  
[\cite{Carlson,Laix}] and has been recently revived in order to address the
issues discussed above [\cite{Spergel}]. The main feature of self-interacting
dark matter (SIDM) is that large self-interacting cross sections lead to a
short mean free path, so that dark matter particles with mean free 
path of the order of the scale length of halos allows for the transfer of 
conductive heat to the halo cores, a quite desirable feature [\cite{Spergel}]. 
Recently performed numerical simulations indicate that strongly 
self-interacting dark matter does indeed lead to better predictions 
concerning satellite 
galaxies [\cite{Hannestad,Moore1,Yoshida,Wandelt}]. However, 
only in presence  
of weak self-interaction [\cite{Burkert}] the core problem might be solved.

The two-body cross section is estimated to be in the range of $\sigma/m \sim 
10^{-24} - 10^{-21}$ cm$^2$/GeV, from a variety of arguments, 
such as the requirement of 
a mean free path between 1 and 1000 kpc [\cite{Spergel}],  the core
expansion time scale to be smaller than the halo age [\cite{Burkert1,Firmani}]
and analysis of cluster ellipticity [\cite{Escude}]. Larger value of 
$\sigma/m \sim  10^{-19} $ cm$^2$/GeV corresponds to the 
best fit to the rotation curve 
of a low surface brightness in  simulations  [\cite{Hannestad}]. 
In our work, we assume for definiteness that the
the cross section is fixed via the requirement  that the mean free path of the
particle in the halo is in the range  $1-1000$ kpc.

\section{A model for self-interacting, non-dissipative CDM}

Many models of physics beyond the Standard Model suggest the existence of new
scalar gauge singlets, {\it e.g.}, in the so-called next-to-minimal
supersymmetric standard model [\cite{NMSSM}]. 
In this section, we provide a simple example for the realization of the 
idea proposed in [\cite{Spergel}] of a self-interacting, non-dissipative cold 
dark matter candidate that is based on an extra gauge singlet, 
$\phi$, coupled to the standard model Higgs boson, $h$, with a Lagrangian 
density given by:

\begin{equation}
{\cal L} = {1 \over 2} (\partial_\mu \phi)^2 - {1 \over 2} m_\phi^2 \phi^2  
- {g \over 4!} \phi^4 + g' v \phi^2 h ~,
\end{equation}
where $g$ is the field $\phi$
self-coupling constant, $m_\phi$ is its mass, $v=246$ GeV is the Higgs 
vacuum expectation value and $g'$ is the coupling between the singlet
$\phi$ and $h$. We assume that the $\phi$ mass does not arise from
spontaneous symmetry breaking since tight
constraints from non-Newtonian forces eliminates this possibility due to the
fact that, in this case, there is a relation among coupling constant, mass and
vacuum expectation value that leads to a tiny scalar self-coupling constant.
In its essential features our self-interacting dark matter model can be 
regarded as a concrete realization of the generic massive scalar field with 
quartic potential discussed in [\cite{Peebles,Goodman}]. We mention that 
a model with features similar to ours has been discussed long ago 
in [\cite{Silveira}]\footnote{We are thankful to A. Zee for pointing 
that out for us.}.

We shall assume that $\phi$ interacts only with $h$ and with
itself. It is completely decoupled for $g' \rightarrow 0$.  
For reasonable values of $g'$, this new scalar would introduce a new, invisible
decay mode for the Higgs boson. This could be an important loophole in the
current attemps to find the Higgs boson at accelerators [\cite{Bij}].
This coupling could, in principle, be relevant for $\phi \phi$ scattering but
we shall be conservative and assume that it is small and neglect 
its contribution.

These particles are non-relativistic, with typical velocities of 
$v \simeq 200 $ km $s^{-1}$. 
Therefore, there is no  dissipation of  energy by, for instance,
creating more particles in reactions like $\phi \phi \rightarrow \phi \phi \phi
\phi$. Only the elastic channel is kinetically accessible and the scattering 
matrix element near threshold ($s \simeq 4\ m_\phi^2$) is given by:
\begin{equation}
{\cal M} (\phi \phi \rightarrow \phi \phi) = i g~~. 
\end{equation}
Near threshold the cross section is given roughly by:
\begin{equation}
\sigma (\phi \phi \rightarrow \phi \phi) \equiv \sigma_{\phi \phi} 
= {g^2 \over 16 \pi s} \simeq {g^2 \over 64 \pi m_\phi^2} ~~.
\label{cross1} 
\end{equation}

We shall derive limits on $m_\phi$ and $g$ by demanding that the mean free path
of the particle $\phi$, $\lambda_\phi$, should be in the interval 
$1~\mbox{kpc} < \lambda_{\phi} < 1~$Mpc. This comes about because, 
if the mean free path  were much greater than about $1~$Mpc, dark 
matter particles would not experience any interaction as they fly through a
halo. On the other hand, if the dark matter mean free path were much
smaller than $1~$kpc, dark matter particles would behave as a collisional  
gas altering substantially the halo structure and evolution. 
Hence, we have:
\begin{equation}
\lambda_{\phi} = {1 \over \sigma_{\phi \phi} n_\phi} 
= {m_\phi \over \sigma_{\phi\phi} \rho^{h}_\phi} ~~, 
\end{equation}
where $n_\phi$ and $\rho_\phi$ are the number and mass density in 
the halo of the $\phi$ particle, respectively. 
Using $\rho^{h}_\phi = 0.4$ GeV/cm$^3$, 
corresponding to the halo density, one finds:
\begin{equation}
\sigma_{\phi \phi} = 2.1 \times 10^{3}~ 
\left({m_\phi \over \gev}\right) 
\left({\lambda_\phi \over \mpc} \right)^{-1}~ \gev^{-2} ~~.
\label{cross2}
\end{equation} 
Equating Eqs. (\ref{cross1}) and (\ref{cross2}) we obtain:
\begin{equation}
m_\phi= 13~g^{2/3}
\left({\lambda_{\phi} \over \mpc} \right)^{1/3} \mev ~~.
\label{mass}
\end{equation}

Demanding the mean free path of the $\phi$
particle to be of order of 1 Mpc implies in the {\it model independent} 
result:
\begin{equation}
{\sigma_{\phi \phi} \over m_\phi} = 8.1 \times 10^{-25} 
\left({\lambda_{\phi} \over \mpc} \right)^{-1} \cm^2/\gev ~~.
\label{SStein}
\end{equation}

Recently, it has been argued, on the basis of  gravitational lensing analysis, 
that the shape of the MS2137 - 23 system is elliptical while 
self-interacting non-dissipative CDM implies that halos are 
spherical [\cite{Escude}]. Furthermore, the limit 

\begin{equation}
{\sigma_{\phi \phi} \over m_\phi} < 10^{-25.5} ~\cm^2/\gev 
\label{Escude}
\end{equation}
arises from that analysis, which is about an order of magnitude smaller 
than (\ref{SStein}). Indeed, gravitational lensing arguments are acknowledged
to be crucial in validating SIDM; however, estimates made in [\cite{Escude}]
were criticized as they rely on a single system and because their intrinsic 
uncertainties actually allow for consistency with SIDM [\cite{Moore1}]. 

In order to  estimate the amount of $\phi$ particles that were produced in the 
early Universe and survived until present,  we  assume that $\phi$ 
particles were mainly produced during reheating after the end of inflation. A 
natural setting to consider this issue is within the framework of 
${\cal N} = 1$ supergravity
inspired inflationary models, where the inflaton sector couples with the 
gauge sector only through the gravitational interaction. Hence, 
the number of $\phi$ particles expressed in terms of the ratio 
$Y_{\phi} \equiv {n_{\phi} \over s_{\gamma}}$, where $s_{\gamma}$ is the 
photonic entropy density, is related with the inflaton ($\chi$) 
abundance after its decay by 

\begin{equation}
Y_{\phi} = {1 \over N} Y_{\chi} ~~, 
\label{inflaton}
\end{equation}
where $N$ is the number of degrees of freedom. Notice that $Y_\phi$ 
is a conserved quantity since $\phi$ does not couple to fermions. 
In the context of ${\cal N} = 1$ Supergravity inflationary models, 
 the upper bound on the 
reheating temperature in order to avoid the gravitino problem (see
[\cite{Bento}]
and references therein), $Y_{\chi}$ is given by the ratio of the reheating 
temperature and the inflaton mass and, for typical models

\begin{equation}
Y_{\chi} = {T_{RH} \over m_{\chi}} = \epsilon~10^{-4} ~~, 
\label{reheating}
\end{equation}
where $\epsilon$ is an order one constant. This estimate allows us to compute
the energy density contribution of $\phi$ particles in terms of the baryonic
density parameter:

\begin{equation}
\Omega_{\phi} = {1 \over N} {T_{RH} \over m_{\chi}} {1 \over \eta_{B}} 
{m_{\phi} \over m_{B}}~\Omega_{B}~~, 
\label{density}
\end{equation}
where $\eta_{B} \simeq 5 \times 10^{-10}$ is the baryon asymmetry of the 
Universe.

Using Eq. (\ref{mass}) and taking $N \simeq 150$, we obtain: 

\begin{equation}
\Omega_{\phi} \simeq 18.5~\epsilon~g^{2/3}
\left({\lambda_{\phi} \over \mpc} \right)^{1/3} \Omega_{B} ~~, 
\label{finaldensity}
\end{equation}
which allows identifying $\phi$ as the cosmological dark matter candidate,
i.e. $\Omega_{\phi} \simeq \Omega_{DM} \lsim 0.3 $ [\cite{Bahcall}],
for $\epsilon \sim 0.5$, $g$ of order one and $\lambda_{\phi}$ of 
about $1~$Mpc.

We have also found that the $\phi$ particle do not generate dangerous
non-Newtonian forces [\cite{us}].

\section{Conclusions}

In this work [\cite{us}], we suggest that a scalar gauge singlet 
coupled with the Higgs
field in such a way as to give origin to an invisible Higgs is 
a suitable candidate
for self-interacting dark matter. This proposal has 
some quite distinct features. 
Firstly, since gauge invariance 
prevents the scalar singlet to couple to fermions, hence 
strategies for directly searching this dark matter candidate must necessarily 
concentrate on the hunt of the Higgs field itself in accelerators. 
Furthermore, in what concerns its astrophysical and cosmological implications,
the main aspects  of our proposal are quite unambiguously expressed 
by Eqs. (\ref{cross2}), (\ref{mass}) and (\ref{finaldensity}). 
Confronting the result of simulations with our candidate for different 
values of the relevant parameters with observations may turn out to 
be crucial for validating our proposal.  
We would like to add that a recent detailed study of our simple model
[\cite{Burgess}] has shown that our abundance computation is valid only 
if the Higgs field is very weakly coupled to
the scalar gauge singlet ($g' < 10^{-8}$), since otherwise the scalar field 
will reach thermal equilibrium [\cite{Burgess}]. The study of phion 
annihilation
processes via exchange of virtual Higgs particles indicates that in 
order to achieve $\Omega_{\phi} h^2 \lsim 0.3$ requires $g' \gsim 2$ 
[\cite{Burgess}] which 
leads to a Higgs decay width that is distinctly different than the 
Standard Model one.

\acknowledgments

R.R. would like to thank PRONEX (CNPq) for providing travel funds
and Grupo de Detectores Avan\c cados of Centro de F\'\i sica Nuclear da 
Universidade de Lisboa under project PRAXIS/P/FIS/10033/98 
for support during his stay in Portugal. L.T. would like 
to acknowledge the financial support 
from Funda\c c\~ao para a Ci\^encia e a Tecnologia (Portugal) 
under the grant PRAXIS XXI /BPD/16354/98 and the project PRAXIS/C/FIS/13196/98.

\thebibliography

\bibitem{us}[1] M.C. Bento, O. Bertolami, 
R. Rosenfeld and L. Teodoro, \PR {\bf D62} (2000), 041302(R).

\bibitem{Bahcall}[2] N. Bahcall, J.P. Ostriker, S. Perlmutter and P.J. Steinhardt,
\SC {\bf 284} (1999) 1481 and references therein.

\bibitem{Mooreetal2}[3] B. Moore, S. Ghigna,  F. Governato, G. Lake, T. Quinn 
and J. Stadel, \AJL {\bf 524} (1999), L19.

\bibitem{Klypin}[4] A.A. Klypin, A.V. Kravtsov, O. Valenzuela and 
F. Prada, \APJ {\bf 522} (1999), 82.

\bibitem{Governato}[5] F. Governato, B. Moore, R. Cen, J. Stadel, 
G. Lake and T. Quinn, {\NAST} {\bf 2} (1997) 91.

\bibitem{Moore}[6] B. Moore, \NAT {\bf 370} (1994) 629.

\bibitem{Flores-Primack}[7] R. Flores and J.R. Primack, \AJ {\bf 427} (1994) 
L1.

\bibitem{Burkert1}[8] A. Burkert, \AJ {\bf 477} (1995) L25.

\bibitem{Blok}[9] W.J.G. de Blok and S.S. McGaugh, \MNRAS {\bf 290} (1997) 533.

\bibitem{Tyson}[10] J.A. Tyson, G.P. Kochanski and I.P Dell'Antonio, 
\AJL {\bf 498} (1998) L107.

\bibitem{Navarro}[11] J. Navarro, C.S. Frenk and S.D.M. White, \AJ {\bf 490} 
(1997) 493.

\bibitem{Ghigna0}[12] S. Ghigna, B. Moore, F. Governato, G. Lake  and 
T. Quinn and J. Stadel, \MNRAS {\bf 300} (1998), 146.

\bibitem{Ghigna}[13] S. Ghigna, B. Moore , F. Governato, G. Lake, T. Quinn, 
J. Stadel, astro-ph/9910166.

\bibitem{Mooreetal}[14] B. Moore, T. Quinn, F. Governato, J. Stadel 
and G. Lake, astro-ph/9903164.

\bibitem{Hogan}[15] C.J. Hogan and J. Dalcanton, astro-ph/0002330.

\bibitem{Carlson}[16] E.D. Carlson, M.E. Machacek and L.J. Hall, 
\AJ {\bf 398} (1992) 43.

\bibitem{Laix}[17] A.A. de Laix, R.J. Scherrer and R.K. Schaefer,
\AJ {\bf 452} (1995) 452.

\bibitem{Spergel}[18] D.N. Spergel and P.J. Steinhardt, 
\PRL {\bf 84} (2000) 3760

\bibitem{Hannestad}[19] S. Hannestad, astro-ph/9912558.

\bibitem{Moore1}[20] B. Moore, S. Gelato, A. Jenkins, 
F.R. Pearce and V. Quilis, astro-ph/0002308.

\bibitem{Yoshida}[21] N. Yoshida, V. Springel, S.D. White and G. Tormen,
astro-ph/0002362.

\bibitem{Wandelt}[22] B.D. Wandelt, R. Dave, G.R. Farrar, P.C. McGuire, 
D.N. Spergel and P.J. Steinhardt, astro-ph/0006344 

\bibitem{Burkert}[23] A. Burkert, astro-ph/0002409.

\bibitem{Firmani}[24] C. Firmani , E. D'Onghia, V. Avila-Reese, 
G. Chincarini and X. Hern\'andes, astro-ph/0002376.

\bibitem{Escude}[25] J. Miralda-Escud\'e, astro-ph/0002050.

\bibitem{NMSSM}[26] See M. Drees, \IJMP {\bf A4} (1989) 3635 and
references therein.

\bibitem{Peebles}[27] P.J. Peebles, astro-ph/0002495.

\bibitem{Goodman}[28] J. Goodman, astro-ph/0003018.

\bibitem{Silveira}[29] V. Silveira and A. Zee, \PL {\bf B161} (1985) 136.

\bibitem{Bij}[30] See T. Binoth and J. J. van der Bij, 
hep-ph/9908256 and references therein.

\bibitem{Bento}[31] M. C. Bento and O. Bertolami, \PL {\bf B365} (1996) 59.

\bibitem{Burgess}[32] C. P. Burgess, M. Pospelov and T. ter Veldhuis, 
{\tt hep-ph/0011335}.

\endthebibliography

\end{document}